**Optimum laser intensity for the production of energetic deuterium ions from laser-cluster interaction**

W. Bang,[1] G. Dyer, H. J. Quevedo, A. C. Bernstein, E. Gaul, J. Rougk, F. Aymond, M. E. Donovan, and T. Ditmire

*Center for High Energy Density Science, C1510, Department of Physics, University of Texas at Austin, Austin, TX, 78712, USA*

**Abstract**

We measured, using Petawatt-level pulses, the average ion energy and neutron yield in high-intensity laser interactions with molecular clusters as a function of laser intensity. The interaction volume over which fusion occurred (1–10 mm$^3$) was larger than previous investigations, owing to the high laser power. Possible effects of prepulses were examined by implementing a pair of plasma mirrors. Our results show an optimum laser intensity for the production of energetic deuterium ions both with and without the use of the plasma mirrors. We measured deuterium plasmas with 14 keV average ion energies, which produced $7.2\times10^6$ and $1.6\times10^7$ neutrons in a single shot with and without plasma mirrors, respectively. The measured neutron yields qualitatively matched the expected yields calculated using a cylindrical plasma model.

---

[1] Author to whom correspondence should be addressed. Electronic mail: dws223@physics.utexas.edu.

## I. INTRODUCTION

There has been much research on the details of the nuclear fusion that results from intense laser interaction with deuterium clusters,[1-6] and the process by which the deuterium ions attain multi-keV kinetic energies has been well explained by the Coulomb explosion model.[7-9] Deuterium clusters are typically 1–10 nm radius assemblies of atoms, bound at liquid density by van der Waals forces. These clusters are produced by forcing cold (80–100 K) deuterium gas under high pressure (~50 bars) through a supersonic nozzle into vacuum. According to the Coulomb explosion model, the electrons in a cluster first absorb the laser pulse energy and the atoms are ionized, a process often referred to as the inner-ionization. After a complete inner-ionization, the cluster consists of positive ions and free electrons. Although the electrons are free to move within the cluster, they are still bound to the cluster. Much more energy is required to free the electrons from the cluster, and this second process is often referred to as the outer-ionization.[10, 11] The electrons gain the energy to escape the cluster through absorption mechanisms such as above-threshold ionization,[12] inverse bremsstrahlung heating,[13-16] resonance absorption,[17-19] nonlinear resonance absorption,[20, 21] and stochastic heating.[22] At high enough laser intensity, almost all of the electrons are removed from the cluster on a time scale short relative to the ion motion, and the remaining highly charged cluster of ions at liquid density promptly explodes by Coulomb repulsion. In cluster fusion experiments, the resultant deuterium ions are energetic enough for collision of those ions to generate DD fusion reactions and produce a burst of 2.45 MeV neutrons. A recent experiment using a 120 J, 170 fs laser pulse reported a fusion yield of $1.6 \times 10^7$ neutrons in a single shot.[23]

The DD fusion cross section rapidly increases with increasing ion energy in the 1–100 keV energy range.[24] Therefore, generating more energetic deuterium ions is crucial to achieving higher fusion yields. In the Coulomb explosion model, this can be done by decreasing the time over which ionization is taking place, defined as the time interval between the start of the inner-ionization and the completion of the outer-ionization. Above all, this ionization time must be short compared to the cluster expansion time, characterized by the time it takes for a cluster to expand to twice its initial size[25]. Optimizing the ionization time experimentally, however, is limited by characteristics of the ultrashort pulsed laser systems, which typically have optical pulses known as "prepulses" that precede the main laser pulse. Picosecond-duration prepulses were experimentally shown to ionize and destroy argon clusters before the arrival of the main pulse, even at laser intensities as low as $\sim 10^{13}$ W/cm$^2$.[26] This effectively decreases the atomic density of the deuterium cluster upon arrival of the main pulse, in turn resulting in a lower kinetic energy Coulomb explosion.

Our previous measurements of average ion energies vs. incident laser intensities using a petawatt-level pulse suggested similar pre-heating of cluster targets by one or more prepulses.[23] In that experiment, deuterium clusters were irradiated by a petawatt laser pulse, and energetic ions (~10 keV) were produced from the laser-cluster interaction. The average kinetic energy of these ions after the Coulomb explosion was measured at different laser intensities using a time-of-flight (TOF) diagnostic, and showed an unexpected decrease at high laser intensities (> $\sim 10^{17}$ W/cm$^2$). Multiple prepulses that could trigger pre-expansion of the clusters were identified in the Texas Petawatt laser (TPW)[27] system, and a one-dimensional (1D) simulation verifies that these prepulses could explain an observed *decrease* in the average ion energy for increasing laser intensities above $1 \times 10^{17}$ W/cm$^2$.[23]

In this paper, we report on experiments on TPW where we varied the level of prepulses in the laser-cluster interaction. We did this by alternatively implementing a pair of plasma mirrors, a proven technique for significantly reducing prepulses while keeping the main pulse intensity at a similar level (~50% in this experiment).[28-31] The neutron yield and the average ion energy were simultaneously measured both with and without[23] plasma mirrors, over a range of laser intensities. The measured neutron yield was compared with the expected yield calculated using the measured ion energy. With plasma mirrors in place, we observed high average ion energies up to 14 keV, but still observed a drop in the average ion energy at the highest laser intensities, similar to what was seen without plasma mirrors.[23]

## II. EXPERIMENTAL SETUP

The TPW is a chirped pulse amplified laser system that can deliver 180 J pulses of 1057 nm wavelength light, with pulse duration down to 130 fs.[27] After the amplified chirped pulse is compressed down to a 130–200 fs pulse, the laser beam continues in vacuum, directed to reflect off an f/40 spherical mirror with a 10 m focal length that focuses the 22 cm diameter beam to a 200 μm diameter focal spot in the target chamber. This focusing geometry creates a much bigger interaction volume than those observed in previous cluster fusion experiments, and higher neutron yields are expected from this increase. We searched for the optimum intensity that produced ions with the highest average kinetic energy by varying the distance between the optical focus and the position of the cluster-producing nozzle, as shown in Fig. 1.

The laser parameters such as the pulse duration, pulse energy, and spatial profile of the laser beam were closely monitored on each shot. When plasma mirrors were not used, the energy of the laser pulse delivered to the target was 102 ±13 J.[23] When double plasma mirrors were used, there was an energy loss of about 50%, so the delivered energy to the target was 55±3 J during the full energy shots. The pulse duration was around 170 fs before the focusing mirror, which was measured by a second order autocorrelator on each shot. Although not directly measured after reflection from the plasma mirrors, we can reasonably assume that the pulse duration remained similar in the plasma mirror case.[28] Laser energy not absorbed from the laser-cluster interaction was attenuated by a known fraction, and directed onto a calibrated energy meter. By comparing this calibrated measured energy with one taken before the focusing mirror, we determined the laser energy absorbed by the cluster jet.

A camera imaged the laser beam at the plane equivalent to the target plane, to represent the actual beam profile on the jet. The average incident laser intensity on the cluster target after the plasma mirrors varied from $10^{15}$ to $10^{17}$ W/cm$^2$, while the laser intensity was $10^{14}$–$10^{18}$ W/cm$^2$ on shots without plasma mirrors. The spatial average of the incident laser intensity was calculated by dividing the pulse energy by the pulse duration and the beam area at the target, all of which were measured from the TPW on-shot diagnostics. The peak laser intensity was up to 3–4 times the average laser intensity in this experiment, and the perimeter of the beam was more intense than the center according to the camera images taken at the equivalent target plane.

For the laser intensity scan, the beam area should be kept similar throughout the entire plasma length. Without the depletion of the incident laser pulse energy, this means that each cluster experiences similar laser intensity throughout the plasma. When the pulse energy is efficiently absorbed by the clusters, we can still examine the collective behavior of the hot ions in terms of the average incident laser intensity. Owing to the large f-number of TPW focusing optics, the Rayleigh length of the TPW laser beam was over 2 cm, which allowed for a scan of the laser intensity by moving the nozzle away from the focus in the setup shown in Fig. 1.

A deuterium cluster jet was produced by cryogenically cooling deuterium gas to 86 K, and forcing it through a pulsed supersonic nozzle at a pressure of 52.5 bars. The conical jet nozzle had a half-angle opening of 5 degrees, an input diameter of 790 μm, and an exit diameter of 5 mm. An XYZ manipulator with 5 cm travel distance in each direction moved the position of the gas jet assembly in vacuum. A series of Rayleigh scattering measurements showed that the average diameter of deuterium clusters was about 16 nm (~100,000 atoms per cluster).

Six plastic scintillation detectors measured the DD fusion neutron yields on each shot, all of which were calibrated using a cluster fusion neutron source from a 20 TW Ti:Sapphire laser prior to this experiment.[32] Four of them used 4.0 cm diameter, 2.0 cm long EJ-232Q scintillators with 105 ps rise time and 700 ps decay time for accurate neutron TOF measurement, and the other two used 4.6 cm diameter, 4.9 cm long EJ-200 scintillators for a higher detection sensitivity. A Photonis XP2020 photomultiplier tube was coupled to each scintillator. Three detectors were located 1.9 m away from the nozzle, and the other three were 4.5 m away to provide a wider detection range. All the detectors were installed at ±90 degrees from the laser propagation direction.

A Faraday cup under vacuum 0.96 m away from the interaction region measured the total number of energetic deuterium ions and their energy distribution function in a time-of-flight (TOF) configuration. This ion TOF diagnostic closely represents the kinetic energies acquired by the exploding clusters at the time of fusion reactions according to a recent experiment[33] that compared the ion temperature during fusion reactions with the average ion energy measured from the TOF diagnostic. An electric field of 80 V/mm was applied over the last 5 mm of this flight path to repel slow electrons that would arrive at the same time as the ions affecting their energy distribution measurement.

Two cameras imaged the ionized region in the cluster jet, one from the side and one from the bottom on each shot for the calculation of the plasma volume. Figures 2(a) and 2(b) show two examples of the side camera images, where the former shows a typical side image of the plasma when the jet was near the laser focus and the latter shows an image when the jet was far away (~10 cm) from the focus. Based on the side and bottom images, the deuterium plasma can be approximated as a cylindrical plasma. We can measure the radius $r$ and length $L$ of the cylindrical plasma from the camera images, and express the volume as $\pi r^2 L$. Therefore, the ion number density becomes $n_D = N_{ion}/(\pi r^2 L)$, where the total number of energetic ions, $N_{ion}$, was estimated by scaling the solid-angle detection of the Faraday cup data to a full $4\pi$ isotropic distribution,[34] $r$ was measured from the laser beam profile at the equivalent target plane, and $L$ was measured from the side image. The measured ion density showed considerable shot-to-shot fluctuations, and varied from $1\times10^{18}$ cm$^{-3}$ to $4\times10^{19}$ cm$^{-3}$ with an average density of $7\times10^{18}$ cm$^{-3}$. Since this is consistent with previously measured value of 7 (±4)$\times10^{18}$ cm$^{-3}$ for similar gas jet conditions,[35] we assumed $n_D=7\times10^{18}$ atoms/cm$^3$ for the subsequent calculations. The plasma diameters of 400 µm and 2 mm in Figs. 2(a) and 2(b), respectively, were consistent with the measured beam profiles.

In order to examine the effect of the TPW prepulses on the cluster fusion experiment, shots were taken using a pair of plasma mirrors to diminish any prepulses, without modifying the system significantly.[29] For this, two uncoated BK7 pieces of glass were inserted about 50 cm before the focus, each with an incidence angle of 45 degrees. Previous studies show that the reflectivity of one plasma mirror reaches as high as 70% when the fluence on it exceeds 50 J/cm$^2$.[28, 29] On the TPW, the s-polarized laser beam fluence on the first mirror was about 100 J/cm$^2$, and the beam fluence on the second mirror was estimated to be about 70 J/cm$^2$. Therefore, we estimate the combined reflectivity of the double plasma mirrors to be 50% for the main pulse, and 1% for prepulses, with a contrast ratio improvement of fifty times.

### III. RESULTS AND ANALYSIS

Figure 3 shows neutron yield (neutrons per shot) as a function of the nozzle distance relative to the laser focus. Each point is an average over all six plastic scintillation detectors on each shot. The open circles indicate data taken without the use of plasma mirrors with delivered laser energies of 102±13 J,[23] where the yield peaked at $1.6 \times 10^7$ neutrons at a focus-target distance of 11 cm. A similar scan performed while using two plasma mirrors is shown as solid triangles in Fig. 3. Here, there was an optimum focus-target distance of 5 cm, achieving a neutron yield of $7.2 \times 10^6$ n/shot using the post-plasma mirror energies of 57 J. The plasma mirrors worked as they were designed according to the energy measurement of the transmitted laser pulse and the images of the plasma. Nevertheless, shots with plasma mirrors did not result in higher neutron yield. We believe this is owing to the decrease in the number of energetic deuterium ions since only 50% of the pulse energy went into the target chamber after the double plasma mirrors.

The energetic deuterium ions measured by TOF in Figs. 4 (a), (b), (c), and (d), taken at four different laser intensities of $2\times10^{14}$ W/cm$^2$, $7\times10^{15}$ W/cm$^2$, $2\times10^{16}$ W/cm$^2$, and $8\times10^{17}$ W/cm$^2$, respectively, showed nearly Maxwellian velocity distributions, which is consistent with previous studies.[23, 34] A strong initial x-ray peak near the time of laser arrival is followed by an energetic (> 1 keV) deuterium ion signal in each figure. As the x-ray peak is relatively prompt, the delay between the start of it and the deuterium ion signal is used to calculate the kinetic energy of ions in the plasma. A dashed red line was added to each figure, which consists of an exponentially rising and decaying curve to approximate the initial x-ray peak, and a Maxwellian fit to the ion TOF data. The latter represents a Maxwellian energy distribution function with $kT$ used as a fitting parameter, with $kT$=2.6 keV, 6.4 keV, 9.1 keV, and 2.3 keV in Figs. 4 (a), (b), (c), and (d), respectively. Previous studies suggest that the observed Maxwellian distribution of hot ions originates not from the thermalization of ions, but from the log-normal size distribution of clusters.[34, 36] This is consistent with the SRIM calculations of the mean free paths: for deuterium ions with energies ranging from 3 keV to 14 keV, the mean free paths vary from 2 mm to 7 mm,[34, 37] which are greater than the plasma radius in this experiment.

Assuming a uniform density of the deuterium ions inside the plasma and throughout the gas jet, an expected neutron yield can be calculated using the ion TOF measurements and a cylindrical plasma model described in this paper and in Ref. [34]. We estimate the DD fusion neutron yield in a cluster fusion as follows:[34]

$$Y \approx \frac{t_d}{2} \int n_D^2 <\sigma_{D(D,n)\,^3He} v>_{kT} dV + N_{ion} \int n_D <\sigma_{D(D,n)\,^3He}>_{kT/2} dl, \qquad (1)$$

where $t_d$ is the disassembly time of the fusion plasma, which can be estimated by the time for a deuterium ion to traverse the plasma, $r/<v>$,[2] with $<v>$ being the mean speed of the hot ions, $n_D$ is the number density of deuterium ions, $<\sigma_{D(D,n)\,^3He} v>_{kT}$ is the fusion reactivity for the D(D, n)$^3$He reactions at an ion temperature of $kT$ defined as two thirds of the average ion energy, $dV$ is the volume element of the fusion plasma, $<\sigma_{D(D,n)\,^3He}>_{kT/2}$ approximates the velocity averaged D(D, n)$^3$He fusion cross-section between hot ions in the plasma and cold atoms or ions in the background gas jet, and *dl* indicates integration over the dimension of the gas jet.

The first term in equation (1), called the beam-beam contribution, accounts for the total DD fusion reactions within the hot plasma, with the integration over the volume, *V,* of the plasma. The second term is the beam-target contribution, which accounts for the fusion reactions between hot deuterium ions originating from the exploding clusters and the cold background deuterium atoms or ions in the gas jet, with the integration over the length, *l*, of the gas jet, where *l* varies from *r* to the radius of the gas jet, *R*. Both contributions in equation (1) can be evaluated using $N_{ion}$ and *kT* measured with the Faraday cup ion TOF data on each shot. For example, the beam-beam contribution can be directly calculated using $<\sigma_{D(D,n)\,^3He} v>_{kT}$, $N_{ion}$, and uniform $n_D$.

Given the velocity distribution *f(v)* of the hot ions, we can calculate the fusion reactivity and the velocity averaged fusion cross section using the known DD fusion cross sections.[24] The observed ion energy distribution closely resembled that from Maxwellian (Fig. 4), so we used the Maxwellian distribution function to calculate $<\sigma_{D(D,n)\,^3He} v>_{kT}$ and $<\sigma_{D(D,n)\,^3He}>_{kT/2}$. The latter cross-section is evaluated at *kT*/2 since the overall ion energy distribution is well fit by a Maxwellian with a temperature of *kT*, and the background atoms or ions can be considered stationary.

Figure 5 compares calculated fusion neutron yields, obtained from equation (1) and the measured TOF data, to experimentally measured yields on shots with and without plasma mirrors, the latter being taken from Ref. [23]. On some of the shots, agreement to within measurement error was not seen, which is not surprising given the simplicity of the cylindrical plasma model: First, the shape of the plasma slightly differs from a perfect cylinder. Also, the model assumes a uniform atomic density throughout the gas jet, as the density profile was not measured. Furthermore, shot-to-shot fluctuations in the ion density were not considered when calculating the expected neutron yields. Nevertheless, the expected fusion yields show good qualitative agreement with the measured yield for data taken with or without the plasma mirrors.

Firing several full energy (100–180 J) shots near the focus without using plasma mirrors, we found that the highest possible intensity laser beam on the cluster target did not produce the most energetic deuterium ions or the greatest number of DD fusion neutrons.[23] On a full energy shot, the TPW reaches $1-3\times10^{18}$ W/cm$^2$ peak laser intensity with f/40 focusing. Yet there were no neutron events when the plasma was produced within ±1 cm from the focus. Instead, the detectors showed very strong x-ray peaks and electromagnetic pulse noise produced from the laser-plasma interaction, which limited the minimum detection range to about $10^4$–$10^5$ n/shot. Only after 50 ns delay beyond the saturating x-ray peak could neutron TOF signals be detected, even for the plastic scintillators with very fast decay constants. Therefore, the estimated neutron yield at the focus is lower than $10^4$ n/shot for shots taken without plasma mirrors.

The scintillation detectors occasionally detected fusion neutrons (yield ~ $10^4$–$10^5$ n/shot) with low energy shots when the gas jet was at the focus. This is unexpected because the neutron yield is known to scale as square of the laser pulse energy as a result of the increase in both the volume and the average ion energy.[8, 25, 38] A neutron yield that initiates for laser pulses with energies as low as 6 J but disappears at higher laser intensities implies a potential role of prepulses. To explore independently the potential role of prepulses, we turn to the intensity scans taken both with and without plasma mirrors. This is based on the hypothesis that if prepulses are destroying deuterium clusters, the yield would return if measurements were taken at the same intensities using plasma mirrors that reduce prepulses by a factor of 50. Measurements of the prepulses using photodiodes verified contrast ratio better than $1\times10^7$ outside a 1 ns time window on TPW. Also, single-shot third order autocorrelation measurements confirmed that the contrast ratio was better than $1\times10^4$ inside the 1 ns time window. The latter prepulse measurements, however, do not rule out the existence of prepulses with intensity lower than $1\times10^{-4}$ of the peak laser intensity. Therefore, it is still possible that the cluster targets were affected by prepulses before the main pulse arrived. It is estimated that a prepulse with intensity as low as ~$1\times10^{11}$ W/cm$^2$ could start destroying the cryogenically cooled deuterium clusters.[39, 40]

The average kinetic energy of deuterium ions was measured for different intensities for shots with and without using plasma mirrors. These measurements are shown as a function of the spatially averaged incident laser intensity in Fig. 6. The solid triangles indicate the average ion energy measured at different intensities using plasma mirrors, whereas the open circles represent the ion energy data without the use of plasma mirrors reported in Ref. [23]. The open squares show low energy shots (~6 J per pulse) taken without using plasma mirrors to investigate the ion energy dependence at a low laser intensity regime. Although the role of prepulses was substantially reduced by the plasma mirrors, the ion energy measurements showed a peaked structure similar to that in Ref. [23]. In both cases, the average ion energy increased as the laser intensity increased, consistent with previous numerical simulations[36, 41] and experiments on deuterium cluster targets.[8, 34, 38] Then, the ion energy peaked at about 14 keV for a laser intensity of $2$–$3\times10^{16}$ W/cm$^2$, and subsequently decreased when clusters were irradiated by higher laser intensities.

## IV. DISCUSSION

The reduction of average ion energy with increasing laser intensity in Fig. 6 is inconsistent with the common description of the cluster ionization process. During the laser-cluster interaction, the average kinetic energy of the electrons on a cluster surface is roughly the ponderomotive potential of the laser field, which increases linearly with the laser intensity. Therefore, more electrons are expected to escape from the cluster as the laser intensity increases. Consequently, higher laser intensities should result in higher cluster charge states, which in turn result in more energetic ions from the subsequent Coulomb explosion. This would be true until the laser intensity becomes high enough to strip all the electrons out of a given cluster, a level referred to as the critical intensity. At laser intensities above the critical intensity, equally energetic ions are expected since the clusters would be still fully outer-ionized.[38] In this experiment, this was not the case, and less energetic ions were produced at laser intensities above the optimum intensity.

Although the exact causes of the ion energy drop at high laser intensities have yet to be identified, we consider two possible explanations. For the case of data taken without plasma mirrors, a 1D simulation reproduced qualitatively similar peaked features assuming a prepulse that arrived 5 ps earlier than the main pulse with an intensity of $1\times10^{-4}$ times its peak intensity. For the simulation, we made the following assumptions.

- A prepulse arrives 5 ps earlier than the main pulse with its peak intensity $10^{-4}$ times smaller than the peak laser intensity of the main pulse. The pulse duration of the main pulse and prepulse are both 200 fs. A deuterium cluster of 500,000 atoms at liquid density is irradiated by a $\lambda$=1.057 µm, Gaussian (temporally), flat-top (spatially) laser pulse.

- Clusters experience Coulomb explosion as they interact with the laser pulse. The ions are not stationary, and start expanding symmetrically as soon as the outer-ionization begins.

- During the laser interaction, an electron at the cluster surface acquires an average kinetic energy equal to the ponderomotive potential, $9.33\times10^{-14}\times I(t)[\text{W/cm}^2]\lambda[\mu m]^2$, where $I(t)$ is the intensity of the laser at time $t$, and $\lambda$ is the wavelength of the laser. If the total energy (electric potential + ponderomotive potential) becomes positive, the electron escapes from the cluster immediately and never comes back.

- The charge state of a cluster at any given time can be calculated by integrating the number of escaped laser-driven electrons over the history of its interaction with the laser.

Figure 7 shows the dependence of the modeled maximum kinetic energy of the deuterium ion as a function of the main pulse peak intensity, which qualitatively agrees with the experimental data. This qualitative agreement suggests that prepulses might have been the cause for the average ion energy drop at higher laser intensities. This, however, needs further verifications, as we saw a similar drop when utilizing plasma mirrors.

A second explanation may be relevant for data obtained both with and without the plasma mirrors. The trend shown in Fig. 6 can be expected if cluster outer-ionization becomes less efficient as the main pulse intensity exceeds some optimum intensity. We think the inverse bremsstrahlung heating followed by a resonant heating can explain the observed trend. The inverse bremsstrahlung heating is the most important heating mechanism during the early stage of the laser-cluster interaction when the cluster can be treated as a cold overdense plasma.[42-44] As the laser intensity increases above an optimum intensity, the ponderomotive potential keeps increasing and the electrons become faster. The inverse bremsstrahlung heating, however, becomes less important because faster electrons are less collisional.[43] At laser intensities higher than $\sim 1 \times 10^{16}$ W/cm$^2$, the electron-ion collision cross-section is negligible, and the laser pulse energy is inefficiently transferred to the electrons via inverse bremsstrahlung heating. Therefore, higher laser intensity beam does not necessarily result in more energetic electrons after the pulse is gone. This could result in incomplete outer-ionization of the larger clusters, which produce the fastest ions and hence contribute most to the fusion yield. Incomplete outer-ionization would lead to less energetic ions produced from the incomplete Coulomb explosion.

Inside the clusters, screening effects are important since the initial local number density of deuterium clusters is well above the critical density. The laser field of TPW, however, does penetrate through the deuterium clusters because the collisionless skin depth (=24 nm) is comparable to the average diameter of our deuterium clusters (=16 nm). Owing to the screening, electrons inside a cluster would experience attenuated laser field, and very efficient inverse bremsstrahlung heating can occur even at laser intensities as high as $10^{16}$ W/cm$^2$.

In a simple 1D model using a fixed atomic density of liquid deuterium and a 150 fs duration square laser pulse with varying peak intensities, we confirmed this trend of an optimum intensity for cluster heating via inverse bremsstrahlung. Since the above argument applies to each individual cluster, one might expect lower laser energy absorption by clusters at higher laser intensities. The drop in the energy absorption fraction at higher laser intensities has been observed previously,[45] and Fig. 2 in Ref. [45] could be understood as a result of the inefficient inverse bremsstrahlung heating inside individual clusters with increasing intensities.

According to our calculations, the inverse bremsstrahlung heating alone can explain the generation of several keV ions from the exploding deuterium clusters. There should be, however, additional heating mechanisms to explain the generation of ~14 keV ions at the optimum laser intensity. Resonant heating is one candidate, which occurs inside a cluster only when it expands, dropping its local density. When it occurs, the electric field inside the cluster becomes large and the electrons can collectively gain high kinetic energy (> 10 keV) from the strong electric field. Since this process is not adiabatic, there is a net gain in the kinetic energy of electrons. This would lead to a complete outer-ionization of large clusters, which can explain the high average kinetic energy of ions observed in this experiment. At higher laser intensities, however, the resonant heating condition is not met within the pulse duration because of slower cluster expansion as a consequence of the inefficient inverse bremsstrahlung heating. This scenario can explain the drop in average ion energy above the optimum laser intensity. In fact, a combination of both effects from prepulses and inefficient cluster heating might cause the trend observed in this experiment.

The cylindrical plasma model showed good agreement between the measured neutron yield and the expected yield in Fig. 5. This supports the validity of the model, and allows for an estimation of the neutron yield accessible with a given system. The ion TOF measurements showed up to 14 keV average deuterium ion energy in this experiment, which was consistent with the transmitted energy measurements showing ~90% absorption by the clusters. The ion TOF data also showed that the hot ions carried less than half of the total laser pulse energy.

Given this information, we can estimate the maximum achievable neutron yields on the TPW with current setup.

TPW can deliver 180 J laser pulse with up to 90 J carried by the energetic deuterium ions. Assuming an ion temperature of 10 keV at the optimum intensity, the total number of ions can be calculated using 90 J = 3/2 $kT \times N_{ion}$. With $kT$=10 keV and $N_{ion}$=3.8×10$^{16}$, the DD fusion neutron yield was estimated using the same model presented in this paper. This sets the highest attainable neutron yield limit of 6.5×10$^7$ n/shot with the current setup on TPW system although we can still improve the neutron yield either by optimizing the laser intensity and the gas jet condition to get hotter ions or by increasing the overall atomic density $n_D$ with a modified supersonic nozzle. Since the ion temperature strongly depends on the cluster size,[36] ion temperatures higher than 10 keV are reachable with bigger deuterium clusters or bigger deuterated methane clusters.[46]

## V. CONCLUSIONS

We investigated the effects of prepulses on the neutron yield and the average ion energy in cluster fusion experiment by implementing a pair of plasma mirrors. After optimization of the laser intensity, the gas jet condition, and the laser pulse duration, we achieved 7.2×10$^6$ neutrons in a single shot using plasma mirrors, where up to 1.6×10$^7$ neutrons were produced in a single shot in Ref. [23]. On many shots with and without plasma mirrors, the laser-cluster interaction successfully generated energetic deuterium ions with average ion energy in the range from 5 keV to 14 keV. The ion energy measured at different intensities with plasma mirrors revealed a similar peaked trend that has been observed without using plasma mirrors.[23] To explain this observed relationship between the average ion energy and the laser intensity, we explored two possibilities. The 1D simulation presented in this paper shows qualitative agreement with the data taken in Ref. [23]. Since similar trend was observed with shots using plasma mirrors, we added a discussion about inverse bremsstrahlung heating becoming less efficient at high laser intensities. The experimentally measured fusion neutron yields were shown to agree with the expected yields calculated using the cylindrical plasma model and the ion TOF data.

Regardless of the use of plasma mirrors, the measurements of average ion energy at different intensities show that there can be an optimum laser intensity above which the ion energy drops for a fixed cluster size. This imposes a limit on the average ion energy feasible in a cluster fusion experiment, and implies that we need to increase the volume of the fusion plasma after accomplishing the desired average ion energy to enhance neutron yield further.

**ACKNOWLEDGMENTS**

WB would like to acknowledge generous support by the Glenn Focht Memorial Fellowship. This work was supported by NNSA Cooperative Agreement DE-FC52-08NA28512 and the DOE Office of Basic Energy Sciences.

**Figures**

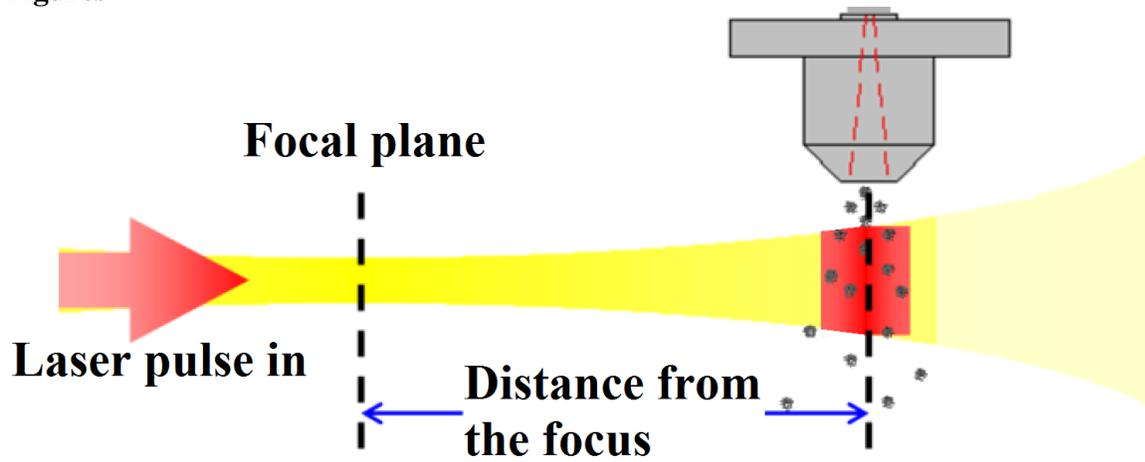

FIG. 1.   (Color online) Schematic diagram for the neutron yield measurement as a function of the distance from the nozzle to the focus

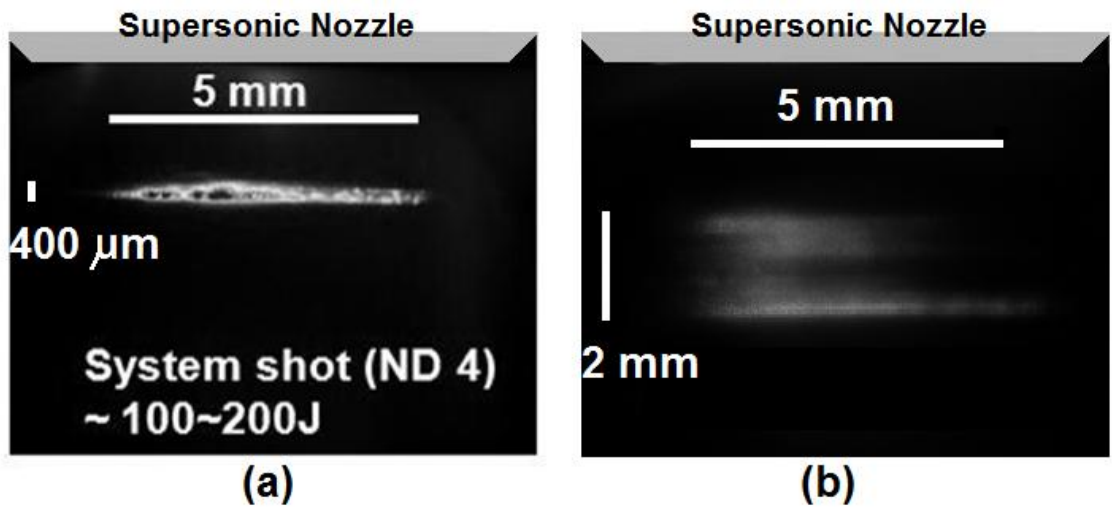

FIG. 2.   Side images of the deuterium plasma on a system shot at (a) laser focus (with laser wavelength suppressing filters) and (b) 10 cm away from the focus.

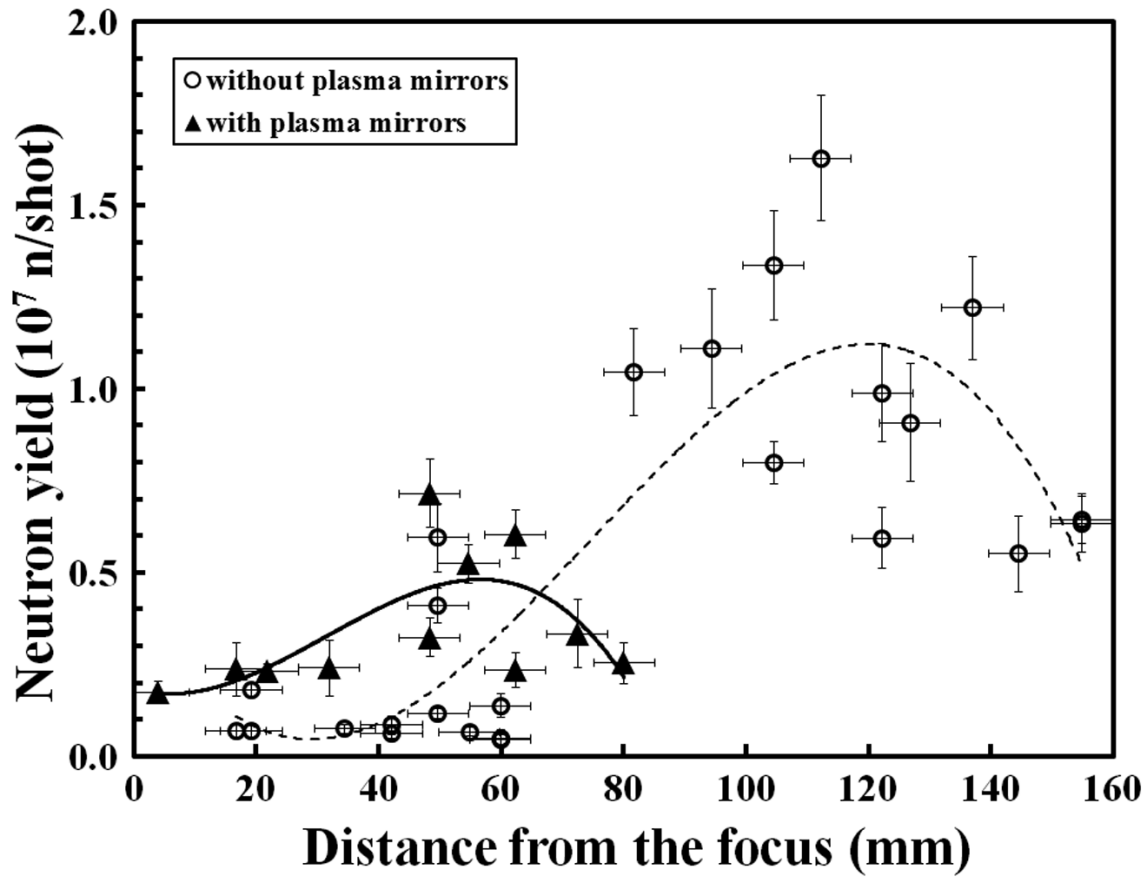

FIG. 3. Neutron yields on shots with (solid triangle) and without (open circle, Ref. [23]) plasma mirrors as a function of the distance from the nozzle to the focus. The solid and dashed lines indicate third order polynomial fits to the data with and without plasma mirrors, respectively, to guide the eye. The vertical error bars represent one standard deviation of the mean.

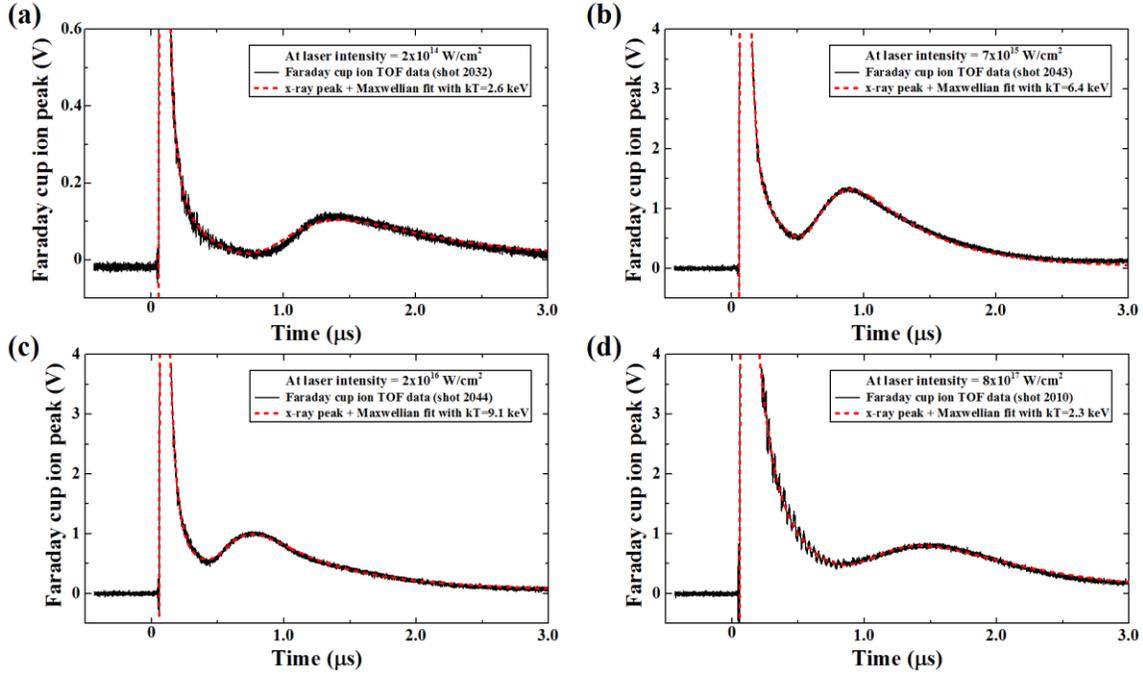

FIG. 4. (Color online) Faraday cup trace showing a strong x-ray peak followed by a deuterium ion signal at four different laser intensities. The average laser intensities were (a) $2\times10^{14}$ W/cm$^2$, (b) $7\times10^{15}$ W/cm$^2$, (c) $2\times10^{16}$ W/cm$^2$, (d) $8\times10^{17}$ W/cm$^2$. The dashed red line in each plot consists of an exponentially increasing and decreasing peak that approximates the initial x-ray peak and a Maxwellian distribution with $kT$, which fits the ion TOF data.

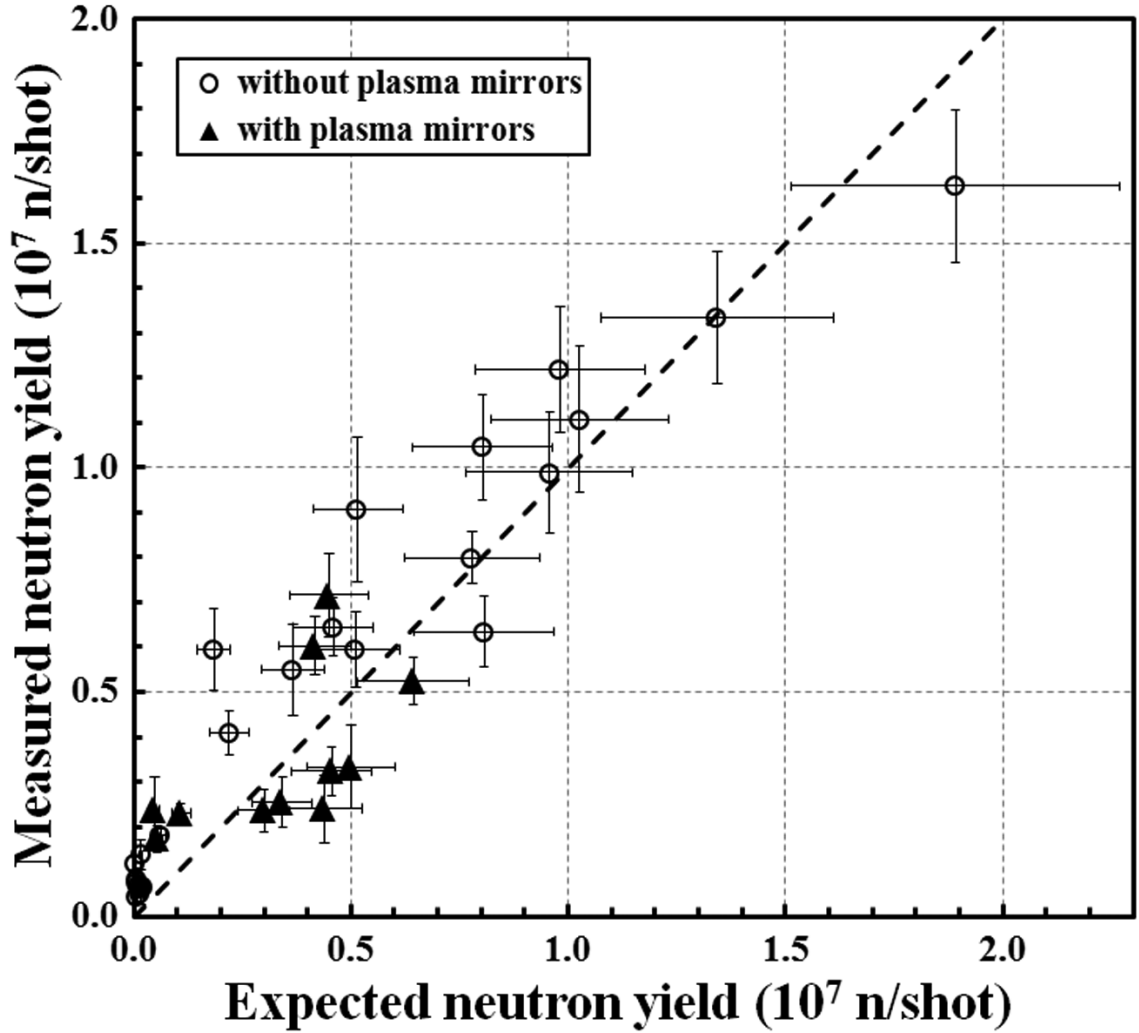

FIG. 5. Neutron yields measured as the average of six scintillation detectors on shots with (solid triangle) and without (open circle, Ref. [23]) plasma mirrors versus the expected neutron yields using equation (1) and the measured TOF data. The vertical error bars indicate one standard deviation of the mean, and the horizontal error bars correspond to 20% error in calculating the expected yield. A dashed line represents where the measured fusion neutron yield matches the expected yield.

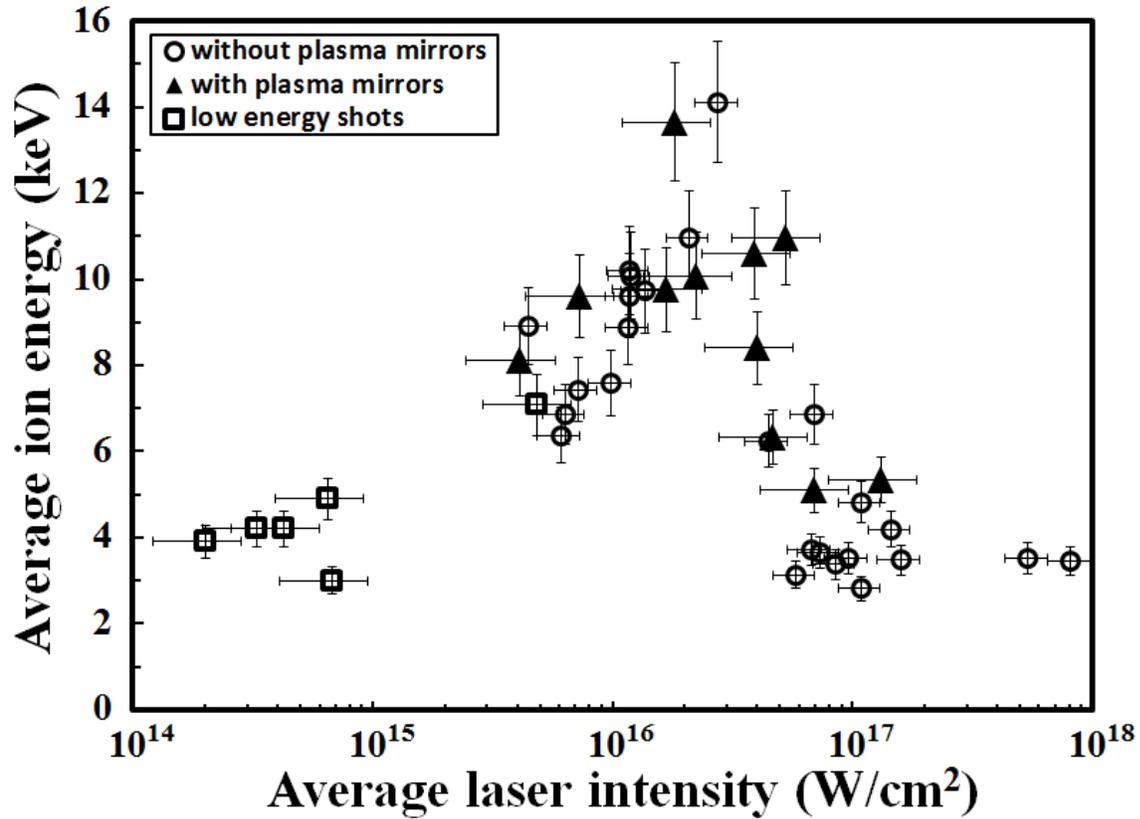

FIG. 6. The average ion energy is plotted for different laser intensities for shots with (solid triangles) and without (open circles, Ref. [23]) plasma mirrors. Low energy shots (open squares, ~6 J per pulse) without plasma mirrors show the dependence at a low laser intensity regime. The error bars indicate 40% uncertainty in measuring the average laser intensity and 10% error in the average ion energy.

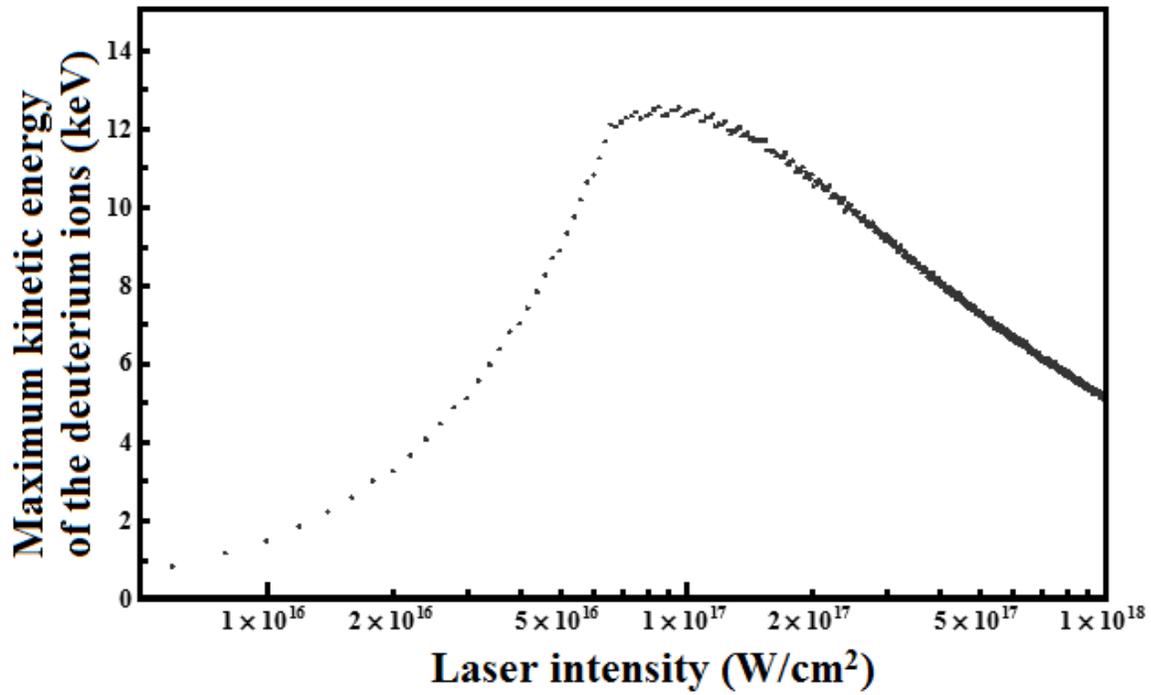

FIG. 7.  A graph showing the calculated maximum kinetic energy of the deuterium ions from an expanding deuterium cluster of 500,000 atoms as a function of the peak intensity of the main laser pulse.